\documentclass[10pt,twocolumn,letterpaper]{article}

\usepackage[pagenumbers]{wacv}

\usepackage[T1]{fontenc}

\usepackage{amsmath}
\usepackage{amssymb}
\usepackage{bm}
\usepackage{float}
\usepackage[utf8]{inputenc}

\definecolor{wacvblue}{rgb}{0.21,0.49,0.74}
\usepackage[breaklinks,colorlinks,allcolors=wacvblue]{hyperref}
\hypersetup{
  pdftitle={GaussianBench: Physics-Fidelity Evaluation for Gaussian Scene Representations},
  pdfauthor={Chukwudalu Dumebi-Kachikwu}
}

\def\wacvPaperID{}
\def\confName{WACV}
\def\confYear{2027}

\title{\textbf{GaussianBench: Physics-Fidelity Evaluation for Gaussian Scene Representations}}

\author{Chukwudalu Dumebi-Kachikwu\\
STAR Laboratories, University of Maryland\\
{\tt\small daludk@terpmail.umd.edu}}

\begin{document}
\maketitle
\begin{abstract}
\begingroup\setlength{\baselineskip}{11.5pt}

3D Gaussian Splatting has rapidly evolved from static scene reconstruction toward dynamic and physics-integrated representations intended not only to reproduce observed motion, but to predict how scenes change under interaction. This shift creates a new evaluation problem: a rollout can look physically plausible while relying on incorrect internal mechanics, and visual agreement with an observed sequence does not establish that the system will respond correctly to a new force, material edit, contact, or thermal intervention. Existing evaluations therefore provide limited evidence about whether the underlying representation actually satisfies the physical behavior it is intended to model.

We introduce \textbf{GaussianBench}, a physics-fidelity evaluation suite for physics-integrated Gaussian scene representations. GaussianBench evaluates physical and representation-level behavior directly through frozen file-based scenes, simulator-independent scorers, and analytical or measured references. The benchmark separately tests conservation, continuum response, heterogeneous-material coupling, Gaussian covariance transport and rendering, thermal phase change, and counterfactual response. Each reference declares its regime of validity, and outcomes distinguish \textsc{pass}, \textsc{fail}, \textsc{na}, and \textsc{invalid}, separating genuine physical failures from unsupported capabilities and invalid comparisons.

We additionally provide GaussianFlesh, a thermomechanical reference entrant in which persistent 3D Gaussians act as both rendering primitives and continuum material points. Its shared-grid UL-MLS-MPM solver supports per-particle constitutive behavior, covariance transport through $\boldsymbol{\Sigma}_p=\mathbf{F}_p\boldsymbol{\Sigma}_{0,p}\mathbf{F}_p^\top$, and persistent thermal, latent-heat, and phase state.

We evaluate seven broad profiles, including six released external systems: PhysGaussian, GaussianFluent, OmniPhysGS, PhysDreamer, Physics3D, and GASP. The shared protocol exposes distinct failure signatures hidden by native visual or task-specific evaluations: systems can reproduce homogeneous continuum response while failing bimaterial coupling, or correctly transport Gaussian covariance while failing to realize the prescribed deformation. Across five-to-ten applicable scenes per external system, GaussianBench separates conservation, continuum, kinematic, representation-state, and validity-regime outcomes, while matched faults and tolerance audits verify that these distinctions arise from the intended contracts. Our results show why physics-integrated Gaussian systems require state-level diagnosis in addition to visually plausible rollouts.

\endgroup
\end{abstract}

\begin{figure*}[t]
    \centering
    \includegraphics[width=0.99\textwidth]{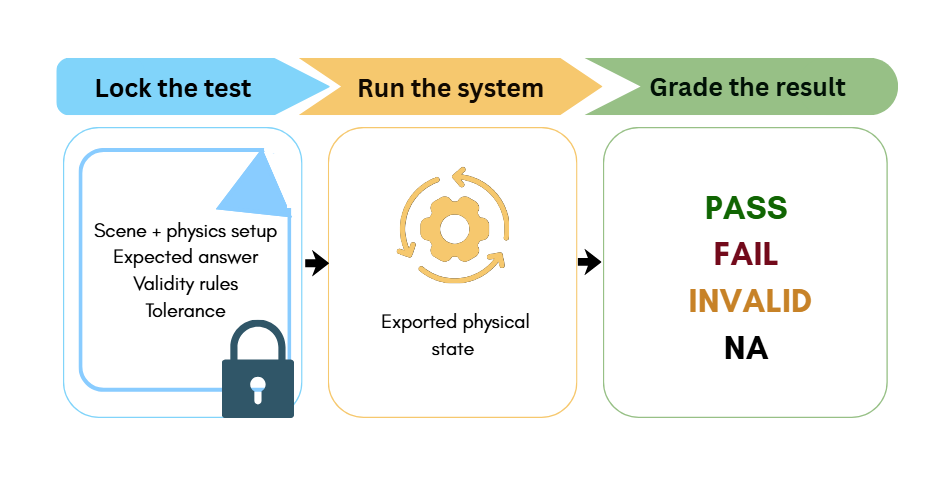}
    \caption{\textbf{GaussianBench in three steps.} We lock each scene's physics setup, reference answer, validity rules, and tolerance; run every released system through a thin adapter that exports its native physical state; and grade the result with simulator-independent scorers that report \textsc{pass}, \textsc{fail}, \textsc{invalid}, or \textsc{na}.}
    \label{fig:benchmark_overview}
\end{figure*}

\section{Introduction}
\label{sec:intro}

Dynamic neural and Gaussian scene representations have made substantial progress in reconstructing time-varying appearance and geometry for novel-view synthesis~\cite{luiten2024dynamic,wu2024fourdimensional}. Physics-integrated Gaussian systems augment that representation with continuum simulation, surface-particle dynamics, or learned material inference~\cite{xie2024physgaussian,feng2025splashing,jiang2024vrgs,ma2025mags,alfonso2026physplat,zhao2025physsplat}. However, if these systems are eventually used to simulate physical phenomena in counterfactual ways, their behavior should be evaluated as carefully as their appearance.

The evaluation gap is now the limiting issue. Published systems are commonly demonstrated on method-specific rendered-motion suites, user studies of motion realism, or bespoke material-edit and fracture examples. For example, PhysGaussian evaluates generated motion on synthetic and captured scenes~\cite{xie2024physgaussian}; PhysDreamer evaluates visual quality and motion realism for action-conditioned rollouts~\cite{zhang2024physdreamer}; OmniPhysGS evaluates material-conditioned interactions~\cite{lin2025omniphysgs}; and GaussianFluent evaluates mixed-material fracture and cutting~\cite{huang2026gaussianfluent}. These are valuable task demonstrations, but they do not create a common measurement of physical state. A rollout can resemble a reference video while failing to conserve momentum, reproduce the correct material response, realize a prescribed deformation, transport Gaussian covariance correctly, or behave correctly under an unseen intervention. In addition, visual evaluation alone cannot reveal which of these physical properties failed, whether the system lacks the capability altogether, or whether the comparison itself is physically valid. GaussianBench supplies that missing evaluation: frozen file-based scenes, pure-function scorers, explicit regimes of validity, and analytical or measured references for mechanics, Gaussian state, rendering, thermomechanics, and material response.

GaussianBench asks a benchmark-first question: which physical and representation-level contracts can a released system demonstrate under a shared, auditable protocol? GaussianFlesh serves as a reference implementation for exercising these contracts and validating the benchmark across mechanical, thermomechanical, and representation-level behavior.

GaussianFlesh treats every Gaussian as a material point with persistent identity. Each particle stores the kinematic, rendering, constitutive, plastic, temperature, latent-heat, and phase state detailed in Sec.~\ref{sec:method}. The same durable memory can store interaction-inferred physical fields or latent material vectors, giving inferred state somewhere to persist and affect later predictions.

A shared UL-MLS-MPM solver advances the cloud. During particle-to-grid (P2G) transfer, each particle reads its own material model ID and dispatches to the appropriate stress law; all particles nevertheless couple through the same grid. This design makes mixed-material objects a property of the particle state rather than a separate scene-graph construction. A single connected object can therefore contain locally stiff, soft, or plastic regions without first being segmented into independently simulated bodies or attached to separate proxy geometry. Those regions exchange momentum through the shared grid, so their interface responds as part of one mechanical system; Sec.~\ref{sec:vandv} tests this coupling against an analytical bimaterial-bar solution. Temperature likewise diffuses on the grid within a thermally simulated body, but the current implementation uses one set of thermal coefficients per body rather than distinct interface coefficients per painted region. This persistent state matters because a physics-integrated system must represent not only an observed trajectory but also how the scene would change under new forces, contacts, temperatures, or material edits.

The demonstrated system supports elastic rubber/jelly, StVK metal with log-strain J2 plasticity and work hardening, Drucker-Prager clay/sand, heat diffusion, continuous thermal softening, latent-heat melting, collisions, and trained 3DGS PLY input.

Our contributions are:
\begin{itemize}
    \item GaussianBench, a file-based physics-fidelity suite with analytical verification of motion, covariance, rendering, thermomechanics, and shared-grid material-interface coupling, plus measured-rubber validation and intervention tests.
    \item An adversarial covariance audit that separates three failure modes: whether the scene deforms correctly, whether Gaussian covariance follows that deformation correctly, and whether the Gaussian state remains healthy rather than collapsing or freezing.
    \item GaussianFlesh, a thermomechanical reference entry with grid heat diffusion, continuous $\mu(T)$ softening, latent heat, and per-particle constitutive dispatch across elastic, J2, Drucker--Prager, and pressure branches.
    \item Cross-system evaluation across seven broad profiles, including six released external systems with five-to-ten applicable scenes each, alongside two separately reported targeted probes.
\end{itemize}

Code, frozen benchmark scenes, adapters, and evaluation reports are publicly available at \url{https://github.com/Dalu1234/GaussianBench}.

\section{Related Work}

\paragraph{Physics simulation.}
Physics-based simulation has enabled the modeling of complex material behavior, including large deformation, plasticity, and phase transitions. The Material Point Method (MPM) has been widely adopted for simulating large-deformation solids and heterogeneous materials, with extensions supporting phase change and varied material behavior~\cite{stomakhin2013snow,stomakhin2014augmented}. Subsequent methods, including the Affine Particle-in-Cell (APIC) method~\cite{jiang2015affine} and Moving Least Squares MPM (MLS-MPM)~\cite{hu2018mlsmpm}, improve particle-grid transfers and computational efficiency. These methods provide a foundation for physics-integrated Gaussian representations, where Gaussian primitives can be coupled to simulators to generate physically driven motion. Our reference implementation, GaussianFlesh, follows this line of research by coupling persistent Gaussian material points through a shared-grid MLS-MPM solver with per-particle constitutive dispatch. By contrast, GaussianBench is independent of the underlying numerical method. Rather than proposing a new discretization or constitutive model, we evaluate whether different simulation approaches meet the same standards for physical accuracy and consistent representation.

\paragraph{Physics-integrated Gaussian representations.}
Recent advances in 3D Gaussian Splatting have extended Gaussian scene representations beyond static reconstruction toward dynamic simulation and physical interaction. Existing approaches employ a variety of physical representations and simulation frameworks. PhysGaussian~\cite{xie2024physgaussian}, OmniPhysGS~\cite{lin2025omniphysgs}, and PhysDreamer~\cite{zhang2024physdreamer} couple Gaussian representations with MPM-based simulation, while Spring-Gaus~\cite{zhong2024springgaus} and PhysTwin~\cite{jiang2025phystwin} employ learned spring-mass representations. Other approaches use PBD/XPBD particles or cages~\cite{feng2025splashing,jiang2024vrgs}, mesh-based representations~\cite{ma2025mags}, or reduced simulation proxies~\cite{alfonso2026physplat}. Relatedly, GASP~\cite{borycki2024gasp} introduces triangle reparameterization for black-box point-based physics, PhysSplat~\cite{zhao2025physsplat} infers physical-property distributions, and GaussianFluent~\cite{huang2026gaussianfluent} addresses mixed-material fracture and cutting.

These approaches demonstrate different capabilities through rendered dynamics, material-conditioned interactions, and task-specific physical simulations. However, their evaluations use different scenes, experimental conditions, and metrics, making it difficult to establish which physical properties each system satisfies under a common protocol. In particular, agreement with observed motion does not by itself establish conservation, constitutive accuracy, or correct Gaussian-state transport. Unlike these approaches, our work focuses on evaluating physical fidelity rather than introducing another simulation architecture. We provide a common set of frozen scenes and simulator-independent scorers that distinguish physical behavior, representation-level correctness, and the validity of each comparison.

\paragraph{Evaluation and verification.}
Physics-integrated Gaussian methods commonly evaluate rendered-video similarity,
perceived motion realism, material-specific demonstrations, predicted geometry,
unseen interactions, or physical-parameter recovery~\cite{xie2024physgaussian,
zhang2024physdreamer,lin2025omniphysgs,huang2026gaussianfluent,
zhong2024springgaus,jiang2025phystwin,li2023pacnerf}. These evaluations provide
valuable evidence, but do not individually establish numerical correctness or
consistency between simulated and rendered state. Drawing on established
verification-and-validation practice~\cite{oberkampf2010verification},
GaussianBench combines analytical code-verification tests, measured-material
validation, Gaussian-state checks, and adversarial scorer audits under a
simulator-independent protocol with declared validity regimes. By separating
failed tests from unsupported capabilities and invalid comparisons, it identifies
which physical and representation-level properties a released system can
demonstrate.

\section{GaussianBench}
\label{sec:vandv}

GaussianBench is a standalone physics-fidelity benchmark for physics-integrated Gaussian splatting systems~\cite{gaussianbench}. Each versioned scene is fixed before entrants are evaluated. Pure-function scorers compare exported state against analytical or measured references and are unit-tested using synthetic trajectories with known outcomes. Held-out variants (unseen versions of benchmark scenes with changed material or geometric parameters and recomputed references) reduce the benefit of instance-specific tuning and test whether a system satisfies the underlying physical contract. The benchmark distinguishes a genuine failure (\textsc{fail}) from a missing capability (\textsc{na}) and a regime violation (\textsc{invalid}). Tier~A grades trajectories, Tier~B grades relational and intervention behavior, and Tier~C grades Gaussian covariance and rendering state. Versioned scene files, manifests, and scorers make every result reproducible and auditable.

\paragraph{GaussianBench Protocols.}
GaussianBench divides evaluation into four complementary protocols. Each asks a
different question about the same exported physical state.

\emph{Protocol A: Does the object obey basic physics over time?}
This protocol follows positions, velocities, and optional mass throughout a
rollout. Conservation scenes check whether an isolated object preserves linear
and angular momentum. Continuum scenes check quantities with known solutions:
beam vibration frequency, stress-wave speed, Poisson contraction, melt-front
motion, and free-fall trajectory. 

\emph{Protocol B: Does the system preserve relationships and respond to changes?}
These scenes test interactions that cannot be judged from one object in
isolation. A bimaterial bar checks whether two regions with different material
parameters develop the predicted regional strains and interface traction.
It does not require different constitutive-law types: each entrant may use an
equivalent native elastic law for both regions, provided it applies the declared
regional parameters and reproduces the assumed small-strain response.
Intervention scenes then change one physical input while holding geometry and initial state fixed, such as gravity, temperature, or material assignment, and compare the new trajectory with the predicted counterfactual. A passing result means the system responds to the changed input rather than replaying a memorized motion.

\emph{Protocol C: Does the Gaussian representation itself remain physically correct?}
This protocol checks whether each Gaussian or native representation carries the
correct physical state as the object deforms. For particle Gaussians, the scorer
independently fits a deformation gradient from the exported positions to measure
whether the requested deformation was realized. It then computes the expected
covariance transport from the entrant's rest covariance and exported deformation
state,
$\boldsymbol{\Sigma}'=\mathbf{F}\boldsymbol{\Sigma}_0\mathbf{F}^{\mathsf T}$,
and compares this result with the covariance exported by the entrant. This
separates errors in the object's motion from errors in Gaussian-state transport.
The transport verdict is therefore conditional on the entrant's exported
$\mathbf F$: it tests whether covariance follows the deformation state that the
system reports, while the separate position-fitted check tests whether the
requested deformation was actually realized.

\emph{Protocol M: Does the constitutive law agree with measured material data?}
The measured-material protocol compares a simulated stress--strain curve with
an independently characterized reference, currently the three-term Yeoh fit to
ASTM~D412 natural-rubber measurements reported by Azarniya and
Rahimi~\cite{azarniya2022rubber,astm_d412}. It is deliberately separate from
analytical code checks: matching a rendered deformation does not establish that
the material law is correct.

These protocols separate four failure modes: wrong motion, wrong
constitutive response, wrong Gaussian-state transport, and an unsupported or
invalid evaluation regime. A \textsc{fail} means the applicable test was run and
missed its reference; \textsc{na} means the released system does not expose the
required capability; and \textsc{invalid} means the prescribed scene regime was
not actually realized.

\paragraph{Adapter validation.}
To ensure that benchmark results reflect the evaluated system rather than adapter mistakes we validate each adapter as part of the measurement pipeline before using its scores: pin the release and harness; verify stable-identity and coordinate round trips; check unit and parameter equality through native conversion functions; measure boundary, gravity, contact, and deformation realization from exported state; trace each graded channel to the field used by the released solver or renderer; and run a matched negative control. A run is marked \textsc{invalid} when the adapter fails to apply a required scene setting, such as gravity, boundary motion, material parameters, or initial state. Missing native state remains \textsc{na}. For covariance-capable entries, we also run a negative control that holds only the covariance fixed. Positions, the native deformation gradient $F$, and all kinematics remain unchanged, so any changed C1 score specifically measures covariance-transport sensitivity. For GASP, which uses a native triangle representation, matched negative controls perturb gravity, state transport, and post-impact recovery within that triangle-state contract. The complete audit matrix is provided in the supplement.

\paragraph{Coverage and reporting policy.}
We distinguish broad profiles from targeted probes. A \emph{broad profile} covers at least five frozen scenes across multiple diagnostic families; a \emph{targeted probe} evaluates a released representation-specific path on only one or two applicable scenes and is not treated as full-system coverage. Seven systems provide broad profiles: GaussianFlesh, PhysGaussian, GaussianFluent, OmniPhysGS, PhysDreamer, Physics3D, and GASP. Spring-Gaus and PhysTwin remain targeted probes because their released implementations do not expose the state or boundary structures required by the broader contracts. 

\begin{table*}[t]
\centering
\scriptsize
\setlength{\tabcolsep}{2.2pt}
\renewcommand{\arraystretch}{1.13}
\begin{tabular}{@{}l l p{0.92in} p{1.17in} p{1.58in} p{2.34in}@{}}
\hline
\textbf{ID} & \textbf{Tier} & \textbf{Reference} & \textbf{Quantity graded} & \textbf{Pass tolerance} & \textbf{Regime of validity} \\
\hline
A1a & A & Conservation & Relative momentum drift (energy reported) & $\Delta p_{\rm rel}<10^{-6}$; $\Delta L_{\rm rel}<10^{-3}$ & Isolated body near its initial location. \\
A1b & A & Conservation & Relative momentum drift (energy reported) & $\Delta p_{\rm rel}<10^{-3}$; $\Delta L_{\rm rel}<10^{-2}$ & Isolated body translating $11.5$\,m; declared domain handling. \\
A2 & A & Euler--Bernoulli & Beam frequency & Relative error $<10\%$ & Slender, linear elastic, small deflection. \\
A3 & A & $c=\sqrt{E/\rho}$ & Wave speed & Relative error $<15\%$ & Thin bar, free lateral surfaces, linear response. \\
A4 & A & Neumann--Stefan & Melt-front prefactor & $R^2>0.98$; error $<15\%$ & 1D conduction, insulated sides, equal phase properties. \\
A5 & A & Poisson law & Lateral contraction & Error $<10\%$; face within $5\%$ & Quasi-static, linear elastic, traction-free sides. \\
A6 & A & Free fall & COM; state health & Position/velocity $<10\%$; volume $<1\%$ & Uniform gravity; no contact in scoring window. \\
B1 & B & Series bar & Regional strain; traction & Each error and jump $<15\%$ & Small strain, equal area, bonded interface. \\
C1 & C & Affine push-forward & Kinematics; transport & Kinematic $<5/10\%$; transport $<0.1/1\%$ & Homogeneous affine interior; boundary band excluded. \\
C2-s & C & Peak shear & Kinematics; transport; health & Same transport limits; declared health limits & Unit volume-preserving shear; regime guarded. \\
C2-m & C & State health & Eigenvalues; condition & Ratio $>10^{-3}$; condition $<10^4$ & Thermal melt capability. \\
C3 & C & Render round trip & PSNR; SSIM & $\geq35$\,dB; $\geq0.97$ & Elastic squash; identical rendering configuration. \\
C4 & C & Measured section & Beam frequency & Relative error $<20\%$ & Irregular captured beam, small deflection. \\
C5 & C & Triangle identity & Edge; area transport & Edge $<1/5\%$; area $<2/10\%$ & Stable triangles; nontrivial translation. \\
C6 & C & Galilean transport & COM path; shape & Straightness $<2\%$; speed var.\ $<5\%$; C5 shape limits & Zero gravity, uniform velocity oblique to the grid. \\
C7 & C & Static equilibrium & Post-settle drift; shape & Drift $<2\%$ body scale; bounded tail shape error & Resting on the system's own released support. \\
C8 & C & Impact recovery & Peak deformation; tail recovery & Peak $\geq5\%$ gate; tail median $<5/10\%$ & Soft-elastic released material; floor impact. \\
M1 & A & ASTM D412 & Cauchy stress & Median error $<15\%$ & Uniaxial, quasi-static, isothermal response. \\
\hline
\end{tabular}
\caption{\textbf{GaussianBench scene inventory and grading contract.} Versioned JSON manifests are normative. Tier A/B consumes trajectories; Tier C consumes exposed representation state and renders.}
\label{tab:scene_inventory}
\end{table*}

\begin{table*}[t]
\centering
\scriptsize
\setlength{\tabcolsep}{4pt}
\begin{tabular}{@{}lccclp{0.39\textwidth}@{}}
\hline
System & Applicable & Families & External & Profile & Observed discrimination \\
\hline
GaussianFlesh & 12 & A/B/C/M & No & Broad & Thermal and bimaterial capability; A1a strict single-precision momentum fails. \\
PhysGaussian~\cite{xie2024physgaussian} & 10 & A/C/M & Yes & Broad & Accurate small-strain response; invalid large-shear realization; covariance transport passes. \\
GaussianFluent~\cite{huang2026gaussianfluent} & 7 & A/C & Yes & Broad & Poisson and covariance pass; conservation, beam, and wave probes fail. \\
OmniPhysGS~\cite{lin2025omniphysgs} & 8 & A/B/C & Yes & Broad & Homogeneous continuum and affine tests pass; bimaterial coupling fails. \\
PhysDreamer~\cite{zhang2024physdreamer} & 8 & A/B/C & Yes & Broad & Continuum and affine tests pass; bimaterial coupling fails. \\
Physics3D~\cite{liu2024physics3d} & 9 & A/C & Yes & Broad & Continuum and covariance contracts of its PhysGaussian-lineage solver pass; strict conservation fails; large shear invalid. \\
GASP~\cite{borycki2024gasp} & 5 & A/C native & Yes & Broad & Free fall, drop transport, Galilean transport, supported settle, and impact recovery all pass on stable triangle state. \\
\hline
Spring-Gaus / PhysTwin & 1--2 each & Native subset & Yes & Targeted & Conservation or free-fall scorer evidence only; hard-coded gravity and no native face structure block the remaining scenes. \\
\hline
\end{tabular}
\caption{\textbf{Coverage accounting used throughout the paper.} Broad profiles support cross-system claims. Targeted probes test simulator and representation independence without being counted as full entries. A/B/C/M denote trajectory/continuum, relational/intervention, representation state, and measured-material families.}
\label{tab:coverage_policy}
\end{table*}

\section{Reference Entrant: GaussianFlesh}
\label{sec:method}

GaussianFlesh is our reference entrant used to test GaussianBench's contracts. It represents a body or captured 3DGS object as persistent Gaussian material points. Each point carries position, velocity, deformation and plastic state, rest and rendered covariance, mass and density, constitutive identifier, and, for thermal scenes, temperature, latent heat, phase, and melt parameters. The complete state table and implementation equations are provided in the supplement.

The production solver is a shared-grid Updated Lagrangian MLS-MPM implementation with APIC-style transfer. Per-particle material identifiers dispatch stress evaluation while all particles scatter to and gather from the same grid. Gaussian appearance is updated from deformation using the covariance push-forward
\begin{equation}
\boldsymbol{\Sigma}_p=\mathbf{F}_p\boldsymbol{\Sigma}_{0,p}\mathbf{F}_p^\top,
\label{eq:covariance}
\end{equation}
so the representation exposes local strain. The reference supports corotated elastic solids, log-strain J2 plasticity, Drucker--Prager response, a weakly elastic compressible molten branch, heat diffusion, latent-heat melting, collisions, and trained PLY input. 

For thermal scenes, heat and phase remain persistent particle state while diffusion is solved on the same grid. Material fields may be painted per particle, allowing heterogeneous regions to exchange momentum through one grid. Parameter inference is outside the present forward-fidelity evaluation, but the persistent state is structured so future inverse models can write inferred material fields without changing the benchmark interface.

\section{GaussianBench Evaluation}
\label{sec:results}

\paragraph{Evaluation scope.}
We compare each system’s published results with results from shared GaussianBench tests. When a test fails, we identify the specific physical contract responsible, check whether the verdict is stable under tolerance changes, and evaluate whether the system responds correctly when a physical input is changed. We
evaluate seven broad profiles, controlled scorer faults, analytical and
measured references, and two targeted probes. Only broad profiles support
cross-system comparisons.

\subsection{Cross-system outcomes under shared tests}

The six external broad entries all report successful demonstrations under their original, method-specific evaluations. GaussianBench does not replace those results or combine their incompatible metrics into a synthetic score. It asks a separate question: which required physical behaviors does each released system reproduce under the same auditable protocol? Four of the six external broad entries fail at least one substantive physical test. Physics3D misses only the strict float32 conservation check, which matters for numerical stability, but does not indicate a broader physics failure. GASP passes its narrower five-scene native-triangle profile.

Table~\ref{tab:native_to_common} pairs each system's published evidence with its
applicable/pass/fail/invalid GaussianBench outcome.

\begin{table}[t]
\centering
\footnotesize
\setlength{\tabcolsep}{3.2pt}
\renewcommand{\arraystretch}{1.08}
\begin{tabular}{@{}lp{0.45\linewidth}c@{}}
\hline
System & Published native evidence (not re-scored) & GB A/P/F/I \\
\hline
PhysGaussian~\cite{xie2024physgaussian} & Synthetic/captured rendered dynamics & $10/7/2/1$ \\
PhysDreamer~\cite{zhang2024physdreamer} & Action-conditioned motion-realism study & $8/5/2/1$ \\
OmniPhysGS~\cite{lin2025omniphysgs} & Material-conditioned interaction suite & $8/5/2/1$ \\
GaussianFluent~\cite{huang2026gaussianfluent} & Mixed-material fracture and cutting suite & $7/2/4/1$ \\
Physics3D~\cite{liu2024physics3d} & Video-guided material optimization demos & $9/7/1/1$ \\
GASP~\cite{borycki2024gasp} & Engine-agnostic Gaussian simulation demos & $5/5/0/0$ \\
\hline
\end{tabular}
\caption{\textbf{Published demonstrations versus GaussianBench results.}
The middle column summarizes what each system demonstrated in its own paper.
The final column reports GaussianBench counts as applicable/pass/fail/invalid.
These are different evaluations, not converted versions of the same metric.
GASP's five applicable scenes use a narrower representation-native profile and
omit conservation, bimaterial, and measured-material families, so its 5/5/0/0
does not imply superiority over systems with broader applicable coverage.}
\label{tab:native_to_common}
\end{table}

\subsection{Attribution of physical failures}
\label{sec:state_results}

\paragraph{Conservation and continuum response.}
These shared tests show that passing a simple single-material simulation does not
guarantee correct overall physics. A system may behave well when the object is
uniform, yet fail to conserve momentum, transmit waves at the correct speed,
bend a beam correctly, or handle multiple materials.
GaussianFluent passes Poisson response but fails conservation, beam, and wave
tests. With grid-coupled prescribed faces, PhysDreamer passes beam frequency,
wave speed, Poisson response, affine realization, and captured frequency, but
fails conservation and bimaterial coupling. OmniPhysGS passes the same
homogeneous continuum and affine probes but also fails the material-interface
test. 

The distinction is clearest in the bimaterial-interface test B1, which joins
two regions with different material parameters and checks both their regional
strains and the traction transmitted across their interface. The largest
regional-strain error is $105.0\%$ for OmniPhysGS and $105.6\%$ for PhysDreamer. At the boundary between the two materials, the force predicted on one side differs from the force predicted on the other by $55.4\%$ for OmniPhysGS and $55.7\%$ for PhysDreamer. Physics3D
produces continuum results close to those of PhysGaussian, consistent with
their shared MPM solver lineage: A2 reaches $4.23\%$, A3 $7.71\%$, A5 $2.17\%$,
and C4 $0.69\%$. It fails only the strict A1a single-precision momentum tier.
No MPM broad entrant passes A1a at its frozen $10^{-6}$ threshold, so we treat
A1a as a numerical-precision diagnostic rather than using it alone as evidence
of model-level physical fidelity. These results show why one overall visual
score or total pass count is not enough: two systems may achieve similar scores
while failing completely different physical tests.

\paragraph{Kinematics versus covariance transport.}
GaussianFlesh passes 10 of the 11 original tests and the added B1
bimaterial-interface test. Its only original failure is A1a, the strict
$10^{-6}$ single-precision momentum check. For the Gaussian-state tests,
GaussianFlesh realizes the prescribed deformation and transports covariance
correctly: C1 (small affine deformation) has a median kinematic error of
$0.020\%$ and native-$F$ transport error of
$3.20\times10^{-6}\%$. The larger-shear C2-s test has
kinematic error $0.075\%$ and transport error
$1.94\times10^{-6}\%$. Thus the benchmark checks two
different things: whether the object moves as prescribed and whether each
Gaussian's covariance follows that motion. 

On the smaller C1 deformation, both PhysGaussian and Physics3D perform well.
For each system, the median kinematic errors are $0.866\%$, and the
median covariance-transport errors are $0.0136\%$. Their independently
computed values coincide after rounding because Physics3D’s released implementation shares an MPM solver lineage with PhysGaussian. The GaussianBench adapter only converts the documented material-parameter units and does not replace or modify the solver. 

Under the harder unit-shear test, each system's median kinematic errors rise to
$45.3\%$, so the systems do not realize the requested deformation
and the scene is marked \textsc{invalid}. The benchmark isolates the failure:
covariances remain healthy and are transported correctly for the deformation
each system actually produces. Thus, the failure is kinematic, not a
covariance-transport error.

C1 and C2-s use different verdict logic for a deliberate reason. Both first
verify that the prescribed boundary faces moved. C1 is a small-deformation code
verification whose graded quantity is whether that drive produces the target
interior affine map, so a wrong interior map is \textsc{fail}. C2-s instead
asks about transport and covariance health at a specified large-shear state;
if the interior never reaches that state, those measurements are not an
evaluation of the declared regime and the row is \textsc{invalid}.When the object’s boundary is moved according to the test through each solver's native grid hook, OmniPhysGS and
PhysDreamer pass C1 at $1.11\%$ median kinematic error. Under C2-s,
their median errors rise to $22.7\%$ rather than the requested unit-shear state and remain
\textsc{invalid}; conditional covariance transport and health still pass.

\paragraph{Heterogeneous response.}
B1 tests an object made from two different materials. In the representative
$h=5$\,mm GaussianFlesh run, the predicted stretching errors are $9.1\%$ and
$7.8\%$ in the two regions, and the force mismatch at their interface is
$1.43\%$. All are below the frozen $15\%$ tolerance. The full three-resolution
sweep is reported in the supplement. This test distinguishes systems that work
for a single uniform material from systems that correctly transfer forces
between different materials.

\paragraph{Representation-native behavior.}
GASP enters five scenes through its released Taichi triangle pipeline and
passes A6, C5, C6 Galilean transport, C7 supported settle, and C8 impact
recovery. Its gravity, half-gravity, and sand-material faults selectively fail
the intended kinematic or recovery metrics. Spring-Gaus exposes momentum drift in A1a/A1b, while PhysTwin
passes native-body A6 free fall. 

\subsection{Robustness to controlled changes}

\paragraph{Adversarial scorer audit.} The audit changes one state component at a time: analytical native-$F$ transport passes, frame-zero covariance under shear fails transport but passes health, and an induced eigenvalue collapse passes transport but fails health.

\begin{center}
\small
\begin{tabular}{lcc}
\toprule
Scorer audit input & Transport & Health \\
\midrule
Exact $F\Sigma_0F^\top$ & Pass & Pass \\
Frozen $\Sigma_0$ under shear & Fail & Pass \\
Induced eigenvalue collapse & Pass & Fail \\
\bottomrule
\end{tabular}
\end{center}

\paragraph{Tolerance sensitivity.}
We re-score all 47 applicable scenes from the six external broad profiles after multiplying every verdict-bearing error budget by $0.5$, $0.75$, $1.25$, and $2.0$; validity gates remain frozen. Doubling the tolerances changes zero verdicts, so none of the external failures or invalid outcomes is created by the nominal cutoff. At $0.75\times$, five borderline passes change; at $0.5\times$, ten passes change. This analysis deliberately targets external entries; it does not establish tolerance robustness for the GaussianFlesh reference entrant. The full per-system sweep and all transitions are machine-readable in the supplement.

\subsection{Response to physical interventions}
\label{sec:interventions}

We use GaussianFlesh to test whether changing a physical input produces the
corresponding predicted change in behavior while geometry and initial state
remain fixed.
In the thermal intervention, we lower the floor temperature from $400\mathrm{K}$ to $370\mathrm{K}$. The classical one-dimensional Stefan melting solution predicts that the melt front should advance at $75.2\%$ of its original rate; GaussianFlesh measures $75.4\%$, corresponding to $0.24\%$ relative error.
In the B1 material-assignment intervention, we stretch a bar divided into left
and right halves and change which half is softer. When both halves have the
same material, physics predicts that the left half should account for
$50.0\%$ of the bar's total extension. Making the left half softer raises its
predicted share to $80.0\%$, while making it stiffer lowers the share to
$20.0\%$. GaussianFlesh measures $49.8\%$, $79.8\%$, and $20.0\%$,
respectively, differing from the predictions by at most $0.23$ percentage
points.

\begin{table}[t]
\centering
\footnotesize
\setlength{\tabcolsep}{4pt}
\begin{tabular}{@{}lrrr@{}}
\hline
Intervention readout & Predicted & Measured & Error \\
\hline
$a_{370}/a_{400}$ melt prefactor & $0.752$ & $0.754$ & $0.24\%$ \\
Uniform soft, $\Delta_L/\Delta$ & $0.500$ & $0.498$ & $0.20$\,pp \\
Soft$\rightarrow$stiff, $\Delta_L/\Delta$ & $0.800$ & $0.798$ & $0.23$\,pp \\
Stiff$\rightarrow$soft, $\Delta_L/\Delta$ & $0.200$ & $0.200$ & $<0.01$\,pp \\
\hline
\end{tabular}
\caption{\textbf{Counterfactual response with fixed geometry.} Only the thermal boundary or material painting changes. ``pp'' denotes percentage points.}
\label{tab:interventions}
\end{table}

\paragraph{Measured material validation.}
Protocol M tests whether a simulated material reproduces a measured
stress--strain curve. We compare the fixed-corotated material model with
natural-rubber tensile measurements collected under ASTM~D412 and represented
by the three-term Yeoh fit of Azarniya and
Rahimi~\cite{azarniya2022rubber,astm_d412}. We first calibrate one parameter,
the small-strain shear modulus, using $\mu=2C_{10}$. We then test whether the
model reproduces the shape of the measured curve up to $150\%$ engineering
strain. Its median error is $5.4\%$, below the frozen $15\%$ tolerance.
PhysGaussian and GaussianFlesh obtain the same result because both use the same
calibrated fixed-corotated constitutive law in this homogeneous test. M1 therefore tests whether a material model calibrated at small strain remains
accurate when the rubber is stretched much further. It does not test automatic
material estimation or differences in how each system advances the simulation
over time. The model slightly underpredicts stress near the start of the curve and overpredicts it at larger strains. So the result applies only to the tested rubber and strain
range, not to all rubber materials.

\subsection{What GaussianBench reveals}

The experiments support three conclusions. First, looking successful in a
system's original demonstration and obeying a specific physical rule are
different claims. A method can produce convincing rendered motion while still
failing tests of conservation, material coupling, deformation, or Gaussian-state
transport. Four of six external broad entries reveal at least one additional
non-precision state-level contract violation under GaussianBench. Second, those discrepancies are not
interchangeable. The benchmark separates conservation, continuum response,
material coupling, kinematic realization, covariance transport, representation
health, and validity-regime violations. Third, the intervention and measured
material protocols test whether the represented state predicts the correct
behavior when conditions change. GaussianBench is therefore most useful as a
diagnostic methodology, not as a single leaderboard score.

\paragraph{Limitations.}
GaussianBench measures declared state contracts, not every aspect of physical
reality.
In addition, it covers a limited set of physical scenarios, and each passing threshold applies only within the conditions defined for that test. Tolerances are also tied to declared regimes, and current
thermal and measured-material references cover only selected materials and
boundary conditions. So the benchmark’s verdicts cannot automatically be generalized beyond the physical conditions for which each test and tolerance were designed. C1 transport is conditioned on exported native $F$ and
does not independently test native $F$ against the position-fitted deformation.

\section{Conclusion}
\label{sec:conclusion}

We introduced GaussianBench, a shared evaluation methodology for testing the
physical behavior and internal Gaussian state that visual and task-specific
evaluations often leave unmeasured. Across six released external systems, four
show at least one non-precision failure despite successful demonstrations in
their original evaluations. These failures span conservation, continuum
response, material-interface coupling, deformation realization, and rendering
state, showing that convincing motion alone does not establish physical
correctness.

GaussianBench's main contribution is diagnosis. It separates whether a system
failed to produce the requested deformation, transported covariance
incorrectly, became numerically unhealthy, or never entered the physical
regime required for a valid test. It also reveals distinctions hidden by a
single visual score or pass count: a system may reproduce a uniform material
while failing at an interface, or transport covariance correctly for a
deformation that it failed to realize. Controlled faults verify that the
scorers detect these failure modes, while tolerance analysis shows that the
headline verdicts are not caused by borderline thresholds. Thermal and material interventions on GaussianFlesh
further demonstrate how the benchmark tests responses to changed physical
inputs.

GaussianFlesh serves as the reference entrant for exercising the full
mechanical, thermomechanical, and Gaussian-state contract, but the primary
contribution is the evaluation framework. As Gaussian splatting moves toward
interactive scenes and world models, visual plausibility will not guarantee
correct behavior under new contacts, forces, or material changes.
GaussianBench makes that difference measurable, attributable, and
reproducible.

\paragraph{Availability.}
The complete GaussianBench and GaussianFlesh code, frozen scenes,
pure scorers, released-system adapters, adversarial scorer tests,
reference formulas, machine-readable reports, and representative
artifacts are publicly available at
\url{https://github.com/Dalu1234/GaussianBench}.

{
    \small
    \bibliographystyle{ieeenat_fullname}
    \bibliography{main}
}

\clearpage
\maketitlesupplementary

\makeatletter
\setlength{\@dblfptop}{0pt}
\setlength{\@dblfpsep}{12pt}
\setlength{\@dblfpbot}{0pt plus 1fil}
\makeatother

\section{Implementation and Reproducibility Details}

This appendix records the constitutive dispatch and stored parameter values used
by the implementation. All stresses below are first Piola--Kirchhoff stresses.
The simulator stores the Lam\'e parameters $(\mu,\lambda)$ directly. For reporting,
we recover the equivalent isotropic parameters as
\begin{equation}
\begin{aligned}
 E&=\frac{\mu(3\lambda+2\mu)}{\lambda+\mu},
 &\nu&=\frac{\lambda}{2(\lambda+\mu)},\\
 \kappa&=\frac{E}{3(1-2\nu)}
 &=\lambda+\frac{2}{3}\mu.
\end{aligned}
 \label{eq:suppl_moduli}
\end{equation}
Equivalently, $\mu=E/[2(1+\nu)]$ and
$\lambda=E\nu/[(1+\nu)(1-2\nu)]$. Values in
Table~\ref{tab:suppl_material_parameters} are derived from the stored
single-precision $(\mu,\lambda)$ values, so the displayed $E$ and $\nu$ include
the small rounding error of those presets.
Table~\ref{tab:suppl_scene_models} maps each reported demonstration to its
production constitutive branch.

\subsection{External Entrant Reproduction}

To make each external evaluation reproducible, we pin every released system to a specific repository commit:
PhysGaussian at \texttt{8339ed6aa2cd5d50e100\allowbreak{}1a254a3d95aea678a956},
GaussianFluent at \texttt{de7e10675b9dce589522\allowbreak{}41087bb15cce9bc3f6ef},
OmniPhysGS at \texttt{5ab915b014467b911581\allowbreak{}b0c6df351713d3114a08}, PhysDreamer at
\texttt{95beb71}, Spring-Gaus at \texttt{62a1bb5}, PhysTwin at
\texttt{81c7187}, Physics3D at \texttt{32d9218}, and GASP at
\texttt{186c9c9}. Each GaussianBench adapter exports a stable particle or
primitive identity together with the system's native positions, velocities,
and, when available, deformation and covariance state.

Any compatibility changes are limited to running the released code in the current software environment. These patches replace deprecated library entry points, make visualization dependencies optional, or repair a released initialization fault. They do not modify the system's force, stress, integration, or covariance-update equations.

Complete machine-readable outputs are provided in
\texttt{results/PhysGaussian\_native\_2026-08-27},
\texttt{results/GaussianFluent\_full},
\texttt{results/OmniPhysGS\_full},
\texttt{results/PhysDreamer\_full},
\texttt{results/Spring-Gaus\_full},
\texttt{results/PhysTwin\_full},
\texttt{results/Physics3D\_broad},
\texttt{results/GASP\_broad}.

PhysGaussian and Physics3D use the same deterministic Warp-MPM kernel,
discretization, timestep policy, and frozen inputs for the shared scenes.
Physics3D's adapter additionally applies its released $E/10^7$ unit convention.
After that conversion, the two independently generated C1 and C2-s reports
differ below the precision printed in the tables; the reported pairs are
median/p95 values, not one value per system.

The Physics3D broad entry runs through the shared bench harness pointed at
its repository. Its released \texttt{compute\_mu\_lam\_from\_E\_nu} stores
Young's modulus in units of $10^7$\,Pa (the native ball configuration uses
$E=0.2$ for $2$\,MPa), so scene moduli are passed as $E/10^7$; passing raw
pascals makes the Lam\'e parameters seven orders of magnitude too stiff,
violates the CFL bound by three orders of magnitude, and reproduces the
solver's illegal-memory-access crash in \texttt{p2g}. The benchmark's
declared corotated law maps to the released pure \texttt{jelly\_base} branch;
the default branch adds a learned Maxwell residual whose zero-modulus case is
guarded, and the persistent \texttt{particle\_F\_N} state is initialized to
identity to match its trial state (the released loader leaves it at zero,
which makes the first constitutive update take the SVD of a singular
tensor). Every \texttt{meta.json} records the mapping, the method and harness
commits, timestep, substep count, and grid.

The GASP broad entry runs its released Taichi pipeline unmodified: the native
teddy pseudomesh vertices are simulated by the released taichi\_elements MPM
engine, GASP's released per-frame triangle size-control rule (threshold
$1.0$) is applied after every step exactly as in its demos, and exports use
the released rescaling inverse. The three new triangle-state scenes
(\texttt{c6\_rigid\_transport\_v1}, \texttt{c7\_supported\_settle\_v1},
\texttt{c8\_impact\_recovery\_v1}) were frozen with synthetic
\textsc{pass}/\textsc{fail}/\textsc{invalid} scorer regression tests,
including each scene's natural controlled fault, before any entrant ran
them. C6's uniform initial velocity is chosen as $D^{-1}\mathbf{d}$ for the
export Jacobian $D$, so the exported center-of-mass direction equals the
scene's declared direction exactly. Matched GASP faults are
\texttt{results/GASP\_half\_gravity\_fault} (A6),
\texttt{results/GASP\_gravity\_left\_on\_fault} (C6, the engine's default
gravity is simply not zeroed), and
\texttt{results/GASP\_sand\_material\_fault} (C8, a released inelastic
material that cannot recover); the Physics3D frozen-covariance C1 fault is
\texttt{results/Physics3D\_frozen\_covariance\_fault}.

OmniPhysGS and PhysDreamer run the frozen particle sets directly through their
released MPM cores. Coordinates are translated, never scaled, into each native
grid and translated back on export. PhysDreamer's one-meter A3 bar uses its
released configurable grid limit of $1.2$\,m so that the bar does not touch the
two-cell safety clip. The initial $1.0$\,m domain produced an artificial
first-frame arrival and was rejected before reporting. Spring-Gaus enters only
A1a/A1b because those references are constitutive-calibration independent;
continuum probes remain NA rather than receiving a benchmark-authored mapping
from $E,\nu$ to learned springs. PhysTwin enters A6 on a deterministic native
particle body using its released $5\times10^{-5}$\,s step. A larger exploratory
step produced nonfinite state and was discarded.

For both OmniPhysGS and PhysDreamer, prescribed faces enter through the
released solver's grid callback before G2P. Exact particle-face targets are
retained only as a secondary position clamp. A particle-only post-step drive
was rejected during adapter validation because it did not transmit the declared
Dirichlet motion reliably into the interior.

\subsection{Uniform Adapter Validation}

Because adapter errors could affect benchmark results, every external adapter is independently validated before scoring. The uniform
protocol checks repository and harness pinning, stable-identity coordinate
round trips, native parameter and unit parity, measured scene realization,
native-state traceability, and a matched negative control. An honest NA means
the release does not expose the required mapping or state and the adapter
leaves that scene unentered rather than synthesizing it.

\begin{table*}[t]
\centering
\scriptsize
\setlength{\tabcolsep}{3.1pt}
\begin{tabular}{@{}lcccccc@{}}
\hline
System & Pin & ID/coords & Units/params & Realization & Native state & Matched control \\
\hline
PhysGaussian & Yes & Yes & Yes & Yes & $F,\Sigma$ & frozen $\Sigma$ \\
GaussianFluent & Yes & Yes & Yes & Yes & $F,\Sigma$ & frozen $\Sigma$ \\
OmniPhysGS & Yes & Yes & Yes & Yes & $F,\Sigma$ & frozen $\Sigma$ \\
PhysDreamer & Yes & Yes & Yes & Yes & $F,\Sigma$ & frozen $\Sigma$ \\
Physics3D & Yes & Yes & $E/10^7$ verified & Yes & $F,\Sigma$ & frozen $\Sigma$; gravity \\
GASP & Yes & triangles & native controls & Yes & triangles; A6 $F$ & gravity; transport; sand \\
Spring-Gaus & Yes & spring points & honest NA & entered scenes & $x,v$ & targeted only \\
PhysTwin & Yes & spring points & honest NA & entered A6 & $x,v$ & targeted only \\
\hline
\end{tabular}
\caption{\textbf{Adapter-validation protocol applied uniformly.} A broad entry must satisfy every applicable check. Unsupported channels remain NA and do not increase coverage.}
\label{tab:adapter_validation}
\end{table*}

For the five covariance-capable external adapters,
\texttt{variants/freeze\_covariance.py} keeps the accepted C1 trajectory and
native $F$ exactly unchanged, but replaces the covariance at every later frame
with the covariance recorded in the initial frame (the undeformed starting
state). Table~\ref{tab:adapter_cov_faults} shows that this change turns the
transport sub-verdict from \textsc{pass} to \textsc{fail} for every adapter,
while the independently fitted kinematic error remains unchanged. Every native
C1 row passes before the fault and fails only its transport sub-verdict after
covariance is frozen.

The C1 and C2-s status rules answer different questions. Adapter validation
first confirms prescribed-face motion for both scenes. C1 then grades the
small-deformation interior affine map itself, so failure to propagate that
motion into the target map is \textsc{fail}. C2-s evaluates transport and
covariance health at a specified large-shear state; if that state is not
realized, the requested regime was never reached and the row is
\textsc{invalid}. With grid-coupled prescribed faces, OmniPhysGS and
PhysDreamer pass C1 at approximately $1.11/1.83\%$ median/p95 kinematic error.
Their C2-s interiors reach $22.7/34.9\%$ error instead of the requested
unit-shear state, so those rows remain \textsc{invalid} while conditional
transport and covariance health pass.
\begin{table}[t]
\centering
\scriptsize
\setlength{\tabcolsep}{6pt}
\resizebox{\columnwidth}{!}{%
\begin{tabular}{@{}lrrc@{}}
\hline
System & Native transport & Frozen-$\Sigma$ & Kinematics \\
\hline
PhysGaussian & $0.0136\%$ & $50.74\%$ & unchanged \\
GaussianFluent & $0.0112\%$ & $51.51\%$ & unchanged \\
OmniPhysGS & $3.56{\times}10^{-6}\%$ & $51.54\%$ & unchanged \\
PhysDreamer & $0.00827\%$ & $51.54\%$ & unchanged \\
Physics3D & $0.0136\%$ & $50.74\%$ & unchanged \\
\hline
\end{tabular}%
}
\caption{\textbf{Matched C1 adapter controls.} Values are median relative covariance-transport error. The fault modifies only covariance. PhysGaussian and Physics3D were scored independently; their values coincide at the displayed precision because the two released systems share the relevant solver lineage.}
\label{tab:adapter_cov_faults}
\end{table}

\subsection{Tolerance Sensitivity}

We test whether the benchmark's conclusions depend on arbitrary numerical
cutoffs for the six external broad profiles. GaussianFlesh is not included in
this sweep, so the analysis supports threshold robustness only for the external
verdicts. Let $e$ be a measured error and $\tau$ its nominal allowed error.
A scene passes when
\[
e \leq \tau.
\]
For a sensitivity multiplier $s$, we change only the allowed error:
\[
e \leq s\tau.
\]
Thus $s=1$ is the nominal rule, $s=0.75$ makes the rule $25\%$ stricter,
$s=0.5$ halves the allowed error, and $s=2$ doubles it. We re-score the
retained native exports under each value of $s$ without rerunning any simulator,
so the physical output is held fixed while only the grading strictness changes.
For metrics where higher is better, we first express the score as an error from
the ideal value of one. For example, SSIM, $R^2$, and direction cosine use
$e=1-m$. PSNR is converted to its implied mean-squared error before scaling.
Covariance-health checks use the corresponding inverse rule: a stricter
multiplier raises the minimum acceptable eigenvalue floor, while a more
forgiving multiplier lowers it. Regime gates remain fixed for every multiplier,
including minimum travel, minimum impact deformation, prescribed-face
realization, frame count, and small-deflection bounds. Therefore, relaxing an
error tolerance cannot turn a scene that failed to realize the prescribed
experiment into a valid pass.

\begin{table}[t]
\centering
\scriptsize
\setlength{\tabcolsep}{4.2pt}
\resizebox{\columnwidth}{!}{%
\begin{tabular}{@{}lccccc@{}}
\hline
System & $0.5\times$ & $0.75\times$ & Nominal & $1.25\times$ & $2\times$ \\
\hline
PhysGaussian & 6/3/1 & 7/2/1 & 7/2/1 & 7/2/1 & 7/2/1 \\
GaussianFluent & 2/4/1 & 2/4/1 & 2/4/1 & 2/4/1 & 2/4/1 \\
OmniPhysGS & 2/5/1 & 3/4/1 & 5/2/1 & 5/2/1 & 5/2/1 \\
PhysDreamer & 2/5/1 & 3/4/1 & 5/2/1 & 5/2/1 & 5/2/1 \\
Physics3D & 5/3/1 & 6/2/1 & 7/1/1 & 7/1/1 & 7/1/1 \\
GASP & 4/1/0 & 5/0/0 & 5/0/0 & 5/0/0 & 5/0/0 \\
\hline
\end{tabular}
}
\caption{\textbf{Tolerance sensitivity as pass/fail/invalid counts.}
Across 47 applicable external scenes, doubling all pass-error budgets changes no
verdict. At $0.75\times$, five nominal passes change; at $0.5\times$, ten
change. Validity gates are fixed.}
\label{tab:tolerance_sensitivity}
\end{table}

At $0.75\times$, five nominal passes change to fail: OmniPhysGS A2/A5,
PhysDreamer A2/A5, and Physics3D A6. At $0.5\times$, PhysGaussian A3,
OmniPhysGS A2/A3/A5, PhysDreamer A2/A3/A5, Physics3D A3/A6, and GASP C7
change from pass to fail. No nominal failure changes to pass even at
$2\times$. The executable analysis, complete per-scene transitions, and
JSON/CSV outputs are released in
\texttt{analysis/tolerance\_sensitivity.py} and
\texttt{results/tolerance\_sensitivity}. All verdict tables can be regenerated
with \texttt{python run\_all.py --results-dir <entry>}.

The frozen A1a threshold is intentionally stringent: no MPM broad entrant
passes its $10^{-6}$ relative-momentum criterion. We therefore report A1a as a
numerical-precision diagnostic and do not use it alone as evidence of a
model-level physical failure. For PhysDreamer A5, the adapter passes
$\nu=0.3$ through the release's native
\texttt{set\_E\_nu\_from\_torch} function and couples the prescribed face
through the native grid callback. The measured Poisson-response error is
$7.79\%$, within the frozen $10\%$ tolerance.
\subsection{Stress and Return Maps}

\paragraph{Fixed-corotated elasticity.}
For $\mathbf{F}^{\mathrm e}=\mathbf{R}\mathbf{S}$ and
$J=\det\mathbf{F}^{\mathrm e}$, the production kernel evaluates
\begin{equation}
 \mathbf{P}=2\mu(\mathbf{F}^{\mathrm e}-\mathbf{R})
 +\lambda(J-1)J(\mathbf{F}^{\mathrm e})^{-\top}.
\end{equation}
The polar factor is corrected to a proper rotation when the SVD reports a
reflection.

\paragraph{Stable Neo-Hookean elasticity.}
For the reference entrant, the stable Neo-Hookean branch used by one region of
B1 follows
Smith et al.~\cite{smith2018stable}. With
$\bar\lambda=\lambda+\mu$, $\gamma=1+\mu/\bar\lambda$, and
$J=\det\mathbf F$, its implemented first Piola stress is
\begin{equation}
 \mathbf P=\mu\mathbf F+\bar\lambda(J-\gamma)J\mathbf F^{-\top}.
\end{equation}
It uses the scene-specified $E$ and $\nu$ through the same Lam\'e conversion as
the other solid branches.

\paragraph{Hencky StVK stress and J2 metal return.}
Let $\mathbf{F}=\mathbf{U}\operatorname{diag}(\boldsymbol{\sigma})\mathbf{V}^\top$
and $\boldsymbol{\epsilon}=\log\boldsymbol{\sigma}$. The principal Kirchhoff
trial stress and the corresponding PK1 stress are
\begin{align}
 \boldsymbol{\tau}^{\mathrm{tr}}
 &=2\mu\boldsymbol{\epsilon}
   +\lambda\operatorname{tr}(\boldsymbol{\epsilon})\mathbf{1},\\
 \mathbf{P}
 &=\mathbf{U}\operatorname{diag}\!\left(
 \frac{2\mu\epsilon_i+\lambda\operatorname{tr}(\boldsymbol{\epsilon})}
 {\sigma_i}\right)\mathbf{V}^\top.
\end{align}
Writing $\widehat{\boldsymbol{\epsilon}}=\operatorname{dev}
(\boldsymbol{\epsilon})$, plastic flow occurs when
$\|\operatorname{dev}(\boldsymbol{\tau}^{\mathrm{tr}})\|_2>\sigma_y$.
The implemented radial return is
\begin{align}
 \Delta\gamma&=\|\widehat{\boldsymbol{\epsilon}}\|_2-
 \frac{\sigma_y}{2\mu},\\
 \boldsymbol{\epsilon}^{\mathrm e}&=\boldsymbol{\epsilon}
 -\frac{\Delta\gamma}{\|\widehat{\boldsymbol{\epsilon}}\|_2}
 \widehat{\boldsymbol{\epsilon}},\\
 \mathbf{F}&\leftarrow\mathbf{U}\operatorname{diag}
 (\exp\boldsymbol{\epsilon}^{\mathrm e})\mathbf{V}^\top,\qquad
 \sigma_y\leftarrow\sigma_y+2\mu\xi\Delta\gamma.
\end{align}
This branch stores the projected elastic deformation directly in $\mathbf{F}$
and resets its separate plastic tensor to identity, matching the production
implementation. Consequently, metal-splat covariance shows the returned
elastic deformation; permanent set remains visible at object scale through
the evolved particle positions.

\paragraph{Drucker--Prager return.}
Clay uses the same Hencky trial stress. With
$\alpha=\sin\varphi/(\sqrt{3}\cos\varphi)$ and
$\widehat{\boldsymbol{\epsilon}}=\operatorname{dev}
(\boldsymbol{\epsilon})$, the production return computes
\begin{equation}
 \Delta\gamma=\|\widehat{\boldsymbol{\epsilon}}\|_2+
 \frac{3\lambda+2\mu}{2\mu}\operatorname{tr}
 (\boldsymbol{\epsilon})\alpha.
\end{equation}
If $\Delta\gamma\leq0$, the trial state is retained. If
$\Delta\gamma>0$ and $\operatorname{tr}(\boldsymbol{\epsilon})>0$, the elastic
singular values are projected to one. Otherwise,
\begin{equation}
\begin{aligned}
 \boldsymbol{\epsilon}^{\mathrm e}&=\boldsymbol{\epsilon}
 -\frac{\Delta\gamma}{\|\widehat{\boldsymbol{\epsilon}}\|_2}
 \widehat{\boldsymbol{\epsilon}},\\
 \mathbf{F}^{\mathrm e}&=\mathbf{U}\operatorname{diag}
 (\exp\boldsymbol{\epsilon}^{\mathrm e})\mathbf{V}^\top,\\
 \mathbf{F}^{\mathrm{pl}}&\leftarrow(\mathbf{F}^{\mathrm e})^{-1}\mathbf{F}.
\end{aligned}
\end{equation}

\paragraph{Molten pressure law.}
After phase transition, singular values are safety-clamped to $[0.2,5]$ to
form $\widetilde{\mathbf{F}}$ and $\widetilde J$. The post-transition branch evaluates
\begin{equation}
\mathbf{P}=\kappa(\widetilde J-1)\widetilde J
 \widetilde{\mathbf{F}}^{-\top}+\mu_{\mathrm{reg}}
 \widetilde{\mathbf{F}}.
\end{equation}
For $\mu_{\mathrm{reg}}>0$, this is a weakly elastic compressible
post-transition material, not a viscosity law or a general free-surface
fluid model.

\paragraph{B1 refinement.} Here, $h$ is the particle spacing, so a smaller
value gives a finer simulation. At the finest resolution ($h=5$\,mm,
$1{,}280$ particles), the two regions have strain errors of $9.1\%$ and
$7.8\%$, and the traction mismatch across their interface is $1.43\%$.
Across the full refinement sweep, decreasing $h$ from $10$ to $6.25$ to
$5$\,mm reduces the traction mismatch from $14.2\%$ to $2.0\%$ to
$1.43\%$. All three resolutions pass the frozen $15\%$ tolerance, and the
improvement with refinement shows that the result is not dependent on one
chosen resolution.

\paragraph{Runtime and external diagnostics.} On one RTX~4070 Laptop GPU, simulation runs at $31.7$ FPS for $5{,}000$ particles, $13.4$ FPS for $40{,}000$, and $2.8$ FPS for a $200{,}000$-Gaussian ficus (0.21, 0.31, and 0.72 GB VRAM). The native GASP and Physics3D diagnostics and their matched fault tests are included as supplementary evidence. These experiments validate the adapters and scorers but do not replace the standardized broad verdict matrix used for cross-system comparison.

\subsection{Implementation equations}
With $\mathbf{d}_{ip}=\mathbf{x}_i-\mathbf{x}_p$ and interpolation weights $w_{ip}$, the grid transfer and deformation update are
\begin{align}
m_i &= \sum_p w_{ip}m_p, &
\\ (m\mathbf{v})_i &= \sum_p w_{ip}m_p(\mathbf{v}_p+\mathbf{C}_p\mathbf{d}_{ip}),\\
\mathbf{f}_i &= -\sum_p V_{0,p}\boldsymbol{\tau}_p\nabla_{\!\mathbf{x}}w_{ip}, &
\mathbf{F}^{n+1}_p &= \Delta\mathbf{F}_p\mathbf{F}^n_p.
\end{align}
The complete constitutive, plasticity, thermal diffusion, latent-heat, and molten-pressure equations are given in the preceding supplementary sections. These define the reference entrant only; GaussianBench scorers import no simulator code.

\section{Reference-Entrant Demonstrations}
\label{sec:qualitative}

\begin{figure}[H]
    \centering
    \includegraphics[width=0.96\textwidth]{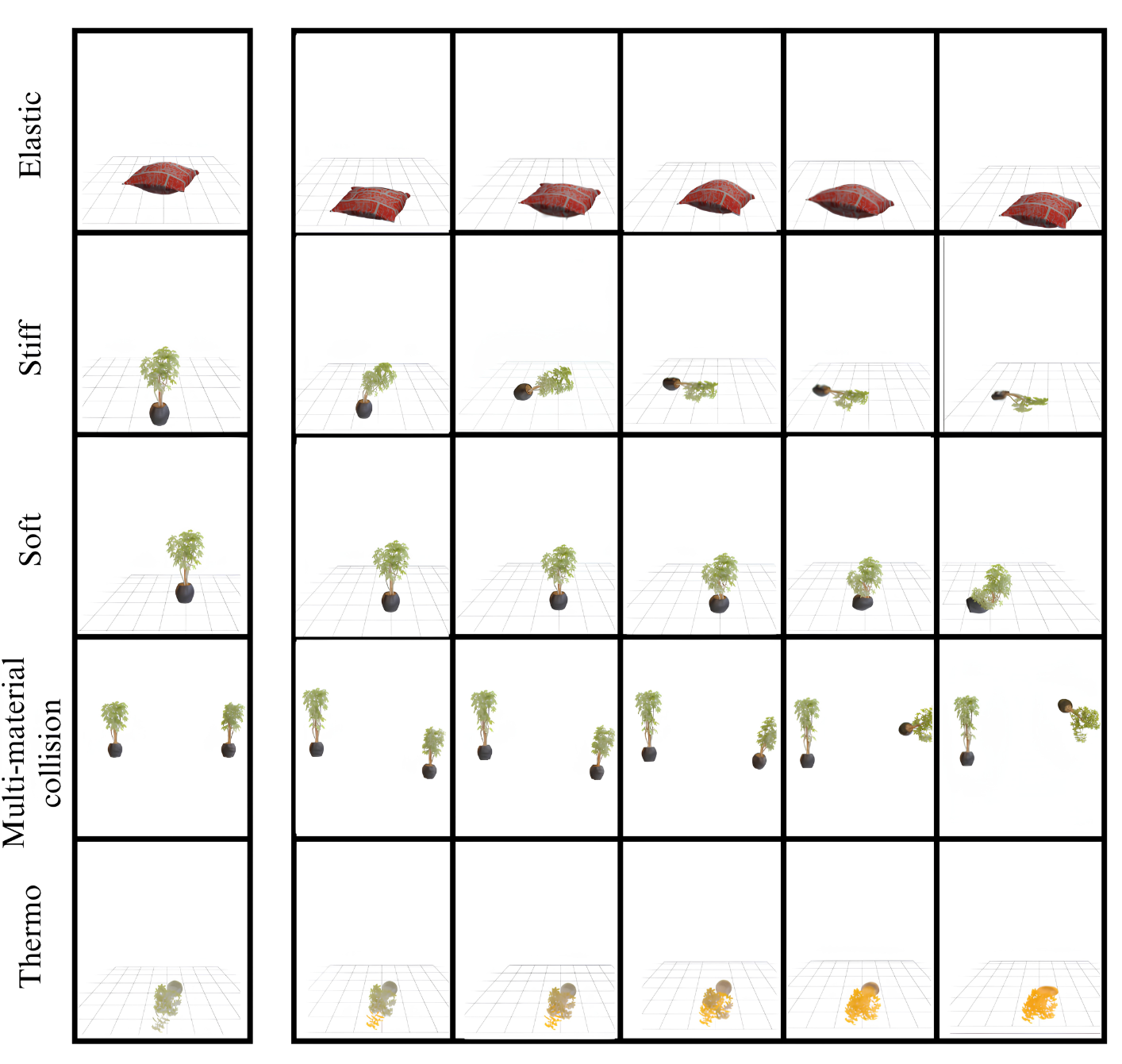}
    \caption{GaussianFlesh reference demonstrations across elastic, stiff, soft, multi-material collision, and thermomechanical regimes. The shared Gaussian substrate carries appearance and persistent physical state.}
    \label{fig:hero_scenes}
\end{figure}

These brief demonstrations show how GaussianFlesh, the reference entrant, realizes selected benchmarked behaviors in rendered Gaussian scenes. The rendered scenes can be seen in Figure~\ref{fig:hero_scenes}.

\paragraph{Material behavior and recovery.}
The reference entry changes per-particle material fields while leaving the representation unchanged. Rubber and jelly use corotated elasticity; metal uses the StVK/log-strain J2 path; clay uses Drucker--Prager; molten material uses the compressible pressure branch. Rubber-like rows in Fig.~\ref{fig:hero_scenes} recover after impact, while stiff/plastic rows retain residual deformation.

\begin{figure}[H]
    \centering
    \begin{minipage}[t]{0.46\textwidth}
        \centering
        \includegraphics[width=\linewidth]{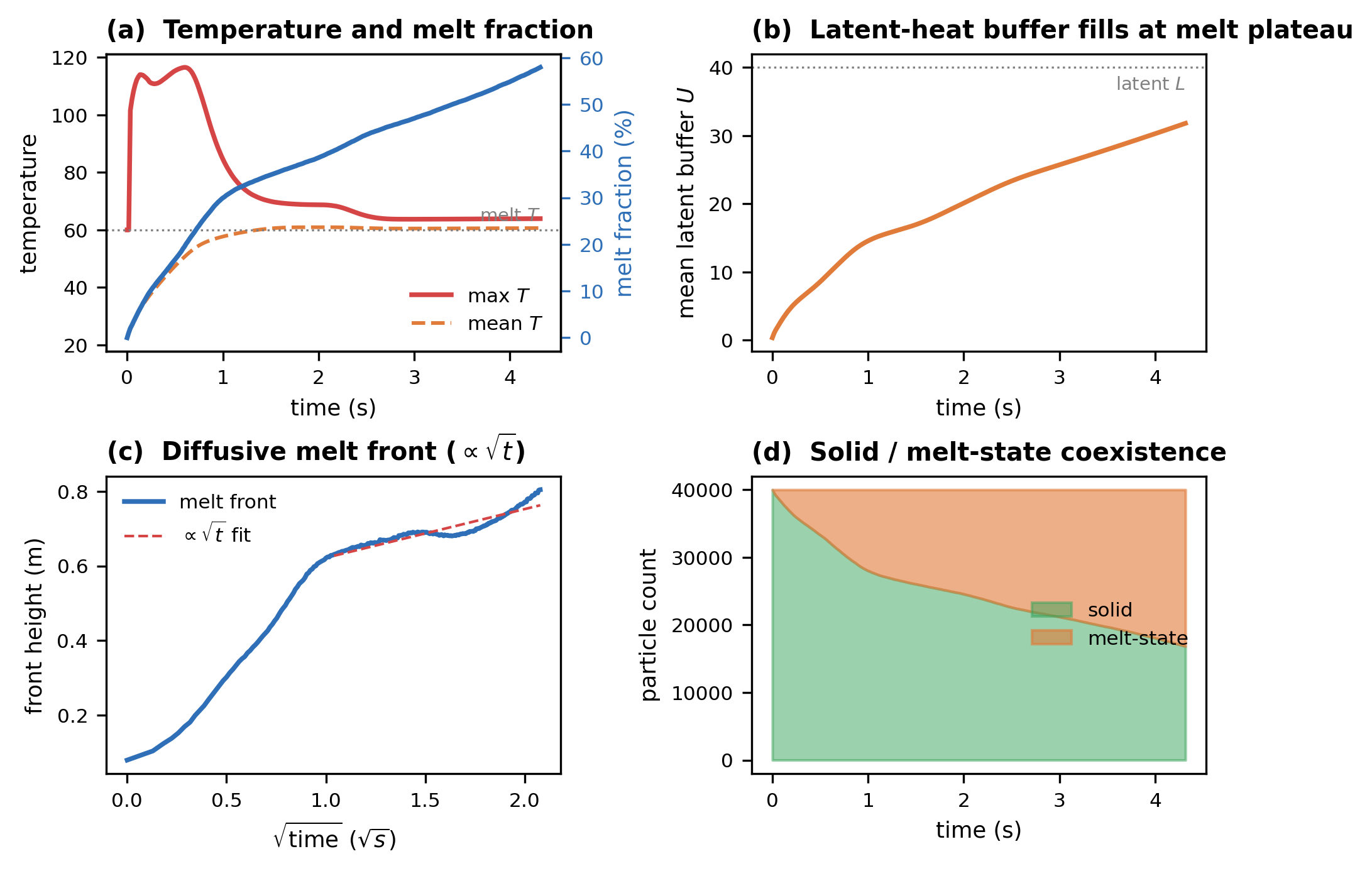}\\[-0.2em]
        \small (a) Heat, latent state, melt fraction, and front position.
    \end{minipage}
    \hfill
    \begin{minipage}[t]{0.38\textwidth}
        \centering
        \includegraphics[width=\linewidth]{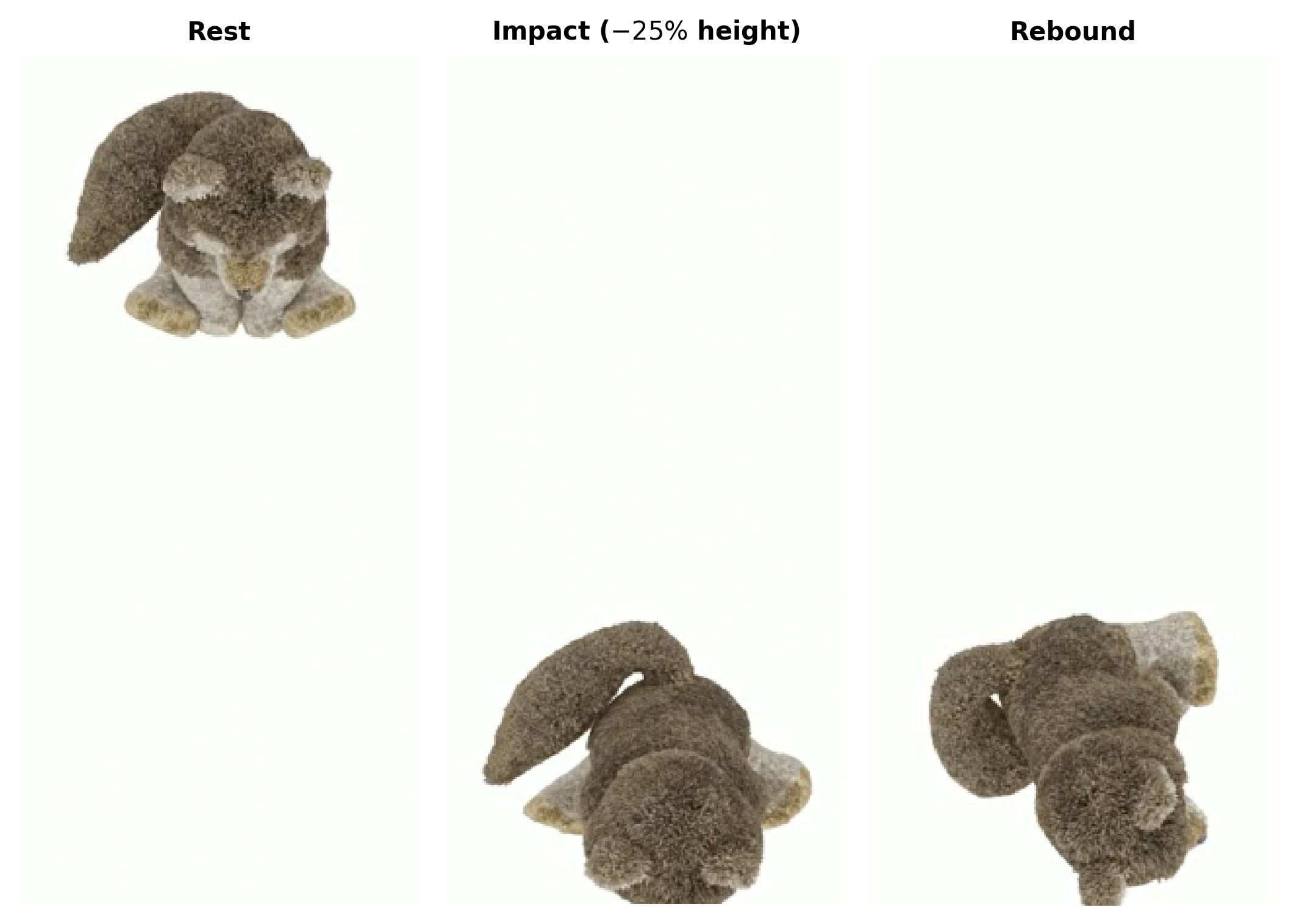}\\[-0.2em]
        \small (b) Elastic round trip on an imported 3DGS capture.
    \end{minipage}
    \caption{Thermomechanical and trained-asset diagnostics for the reference
    entrant: melting-state evolution (left) and deformation recovery of a
    captured plush object (right).}
    \label{fig:thermal}
    \label{fig:roundtrip}
\end{figure}

\paragraph{Thermomechanical melting.}
The coupled path combines grid heat diffusion, continuous $\mu(T)$ softening, latent heat, and phase transition. Figure~\ref{fig:thermal} logs temperature, melt fraction, latent buffer, front position, and phase counts.

\paragraph{Trained 3DGS assets.}
GaussianFlesh loads trained PLY point clouds and reuses learned covariance, opacity, and spherical-harmonic appearance. For a captured plush object, we follow PhysGaussian's internal-filling procedure~\cite{xie2024physgaussian}. After $25\%$ impact compression, the object rebounds to within $0.8\%$ of the bounding-box diagonal, with $29$\,dB PSNR / $0.98$ SSIM~\cite{wang2004ssim} after rigid alignment and no invalid covariances. This texture-sensitive diagnostic is distinct from procedural C3 and remains below C3's $35$\,dB threshold.

\clearpage
\onecolumn
\raggedbottom

\subsection{Complete External System--Scene Results}

Tables~\ref{tab:external_mechanics_matrix} and
\ref{tab:external_state_matrix} expand the aggregate counts in the main paper
into per-scene outcomes for every external broad entry. Each applicable cell
reports the verdict and the primary error used to interpret it; a dash denotes
\textsc{na}. Full metric vectors, references, and tolerances remain in the
released \texttt{report.json} files. These tables report retained results and
do not introduce a new aggregate score.

\begin{table}[H]
\centering
\scriptsize
\setlength{\tabcolsep}{3.6pt}
\renewcommand{\arraystretch}{1.35}
\resizebox{\textwidth}{!}{%
\begin{tabular}{@{}lcccccccc@{}}
\hline
System & A1a & A1b & A2 & A3 & A5 & A6 & B1 & M1 \\
\hline
PhysGaussian & F $8.95{\times}10^{-6}$ & P $1.02{\times}10^{-6}$ & P $4.23\%$ & P $7.71\%$ & P $2.17\%$ & -- & -- & P $5.40\%$ \\
GaussianFluent & F $1.73{\times}10^{-2}$ & F $3.04{\times}10^{-3}$ & F $76.9\%$ & F no arrival & P $0.263\%$ & -- & -- & -- \\
OmniPhysGS & F $4.16{\times}10^{-6}$ & -- & P $8.06\%$ & P $9.04\%$ & P $8.18\%$ & -- & F $105\%$ & -- \\
PhysDreamer & F $2.35{\times}10^{-5}$ & -- & P $8.61\%$ & P $9.04\%$ & P $7.79\%$ & -- & F $106\%$ & -- \\
Physics3D & F $8.56{\times}10^{-6}$ & P $1.00{\times}10^{-6}$ & P $4.23\%$ & P $7.71\%$ & P $2.17\%$ & P $7.60\%$ & -- & -- \\
GASP & -- & -- & -- & -- & -- & P $2.34\%$ & -- & -- \\
\hline
\end{tabular}%
}
\caption{\textbf{Complete external mechanics results.} P/F/I denote
\textsc{pass}/\textsc{fail}/\textsc{invalid}. A1a/A1b report relative linear
momentum drift; A2/A3/A5/M1 report relative error; A6 reports the larger of the
COM position and velocity errors. B1 reports the largest regional-strain or
interface-traction error. A2 invalidity is shown by maximum tip deflection.
A1a is reported as a strict float32 precision diagnostic.}
\label{tab:external_mechanics_matrix}
\end{table}

\begin{table}[H]
\centering
\scriptsize
\setlength{\tabcolsep}{4.2pt}
\renewcommand{\arraystretch}{1.35}
\resizebox{\textwidth}{!}{%
\begin{tabular}{@{}lcccccccc@{}}
\hline
System & C1 kin./tr. & C2-s kin./tr. & C3 & C4 & C5 & C6 & C7 & C8 \\
\hline
PhysGaussian & P $0.866/0.0136\%$ & I $45.3/0.0747\%$ & F $31.63$ dB & P $0.691\%$ & -- & -- & -- & -- \\
GaussianFluent & P $0.864/0.0112\%$ & I $12.4/0.0815\%$ & -- & -- & -- & -- & -- & -- \\
OmniPhysGS & P $1.11/3.56{\times}10^{-6}\%$ & I $22.7/4.01{\times}10^{-6}\%$ & -- & P $6.82\%$ & -- & -- & -- & -- \\
PhysDreamer & P $1.11/0.00827\%$ & I $22.7/0.0265\%$ & -- & P $7.45\%$ & -- & -- & -- & -- \\
Physics3D & P $0.866/0.0136\%$ & I $45.3/0.0747\%$ & -- & P $0.691\%$ & -- & -- & -- & -- \\
GASP & -- & -- & -- & -- & P $0.0302\%$ & P $0.000848\%$ & P $1.41\%$ & P $1.05/0.912\%$ \\
\hline
\end{tabular}%
}
\caption{\textbf{Complete external representation-state results.} C1 and C2-s
report median kinematic/transport error; C3 reports PSNR; C4 reports frequency
error (or invalidating tip deflection); C5 reports area-error p95; C6 reports
path straightness; C7 reports post-settle COM drift; and C8 reports tail
edge/area recovery error. C2-s is invalid when the prescribed shear is not
realized, even though the conditional transport error passes. C1 instead
grades realization of its smaller prescribed affine map after boundary-face
motion has been verified.}
\label{tab:external_state_matrix}
\end{table}

\begin{table}[H]
\centering
\small
\setlength{\tabcolsep}{6pt}
\renewcommand{\arraystretch}{1.45}
\begin{tabular}{@{}lll@{}}
\hline
\textbf{Scene or demonstration} & \textbf{Material assignment} & \textbf{Constitutive path} \\
\hline
Elastic drop and recovery & Rubber or jelly & Fixed-corotated elasticity \\
Metal impact and permanent set & Metal & Hencky StVK stress with log-strain J2 return \\
Clay deformation & Clay & Hencky stress with Drucker--Prager return \\
Mixed connected objects & Region-painted presets & Per-particle dispatch among the above laws \\
Hot-floor melting & Wax solid $\rightarrow$ molten wax & Corotated solid $\rightarrow$ weakly elastic compressible melt \\
Imported ficus, pillow, and plush captures & Assigned rubber or jelly & Corotated physics; trained opacity and SH retained \\
GaussianBench solid scenes & Scene-specified $E$, $\nu$, and $\rho$ & Fixed-corotated elasticity \\
GaussianBench B1 bimaterial bar & Region-painted $E$; entrant-native law mapping & Small-strain-equivalent elastic interface response \\
\hline
\end{tabular}
\caption{Scene-to-constitutive-model mapping. A material label selects particle
fields, not a different solver. Mixed regions scatter to the same UL-MLS-MPM
grid and dispatch their stress law per particle.}
\label{tab:suppl_scene_models}
\end{table}

\begin{table}[H]
\centering
\small
\setlength{\tabcolsep}{5pt}
\renewcommand{\arraystretch}{1.45}
\begin{tabular}{@{}lrrrrrr@{}}
\hline
\textbf{Preset} & $E$ (Pa) & $\nu$ & $\mu$ (Pa) & $\lambda$ (Pa) & $\kappa$ (Pa) & $\rho$ (kg/m$^3$) \\
\hline
Rubber & $1.000{\times}10^6$ & $0.4499$ & $3.45{\times}10^5$ & $3.10{\times}10^6$ & $3.33{\times}10^6$ & $1200$ \\
Jelly & $4.988{\times}10^4$ & $0.4501$ & $1.72{\times}10^4$ & $1.55{\times}10^5$ & $1.665{\times}10^5$ & $1000$ \\
Metal & $1.856{\times}10^8$ & $0.2999$ & $7.14{\times}10^7$ & $1.07{\times}10^8$ & $1.546{\times}10^8$ & $7800$ \\
Clay & $1.000{\times}10^5$ & $0.4499$ & $3.45{\times}10^4$ & $3.10{\times}10^5$ & $3.33{\times}10^5$ & $1800$ \\
Flesh / bread & $5.011{\times}10^4$ & $0.3998$ & $1.79{\times}10^4$ & $7.14{\times}10^4$ & $8.333{\times}10^4$ & $1050$ / $400$ \\
Wax (solid) & $9.722{\times}10^4$ & $0.4090$ & $3.45{\times}10^4$ & $1.55{\times}10^5$ & $1.78{\times}10^5$ & $900$ \\
Wax (molten) & -- & -- & $3.00{\times}10^3$ & -- & $4.00{\times}10^5$ & $900$ \\
\hline
\end{tabular}
\caption{Constitutive parameters stored by the named presets. For solids,
$\kappa$ is the derived isotropic bulk modulus in Eq.~\eqref{eq:suppl_moduli};
the code evaluates stress using $(\mu,\lambda)$. The molten-wax row instead
lists the pressure-law bulk modulus and residual shear elasticity used after
phase transition. Metal additionally uses initial yield stress
$\sigma_y=5.0{\times}10^7$ Pa and hardening coefficient $\xi=10$; clay uses
friction angle $\varphi=30^\circ$.}
\label{tab:suppl_material_parameters}
\end{table}

\begin{figure}[H]
    \centering
    \includegraphics[width=0.68\textwidth]{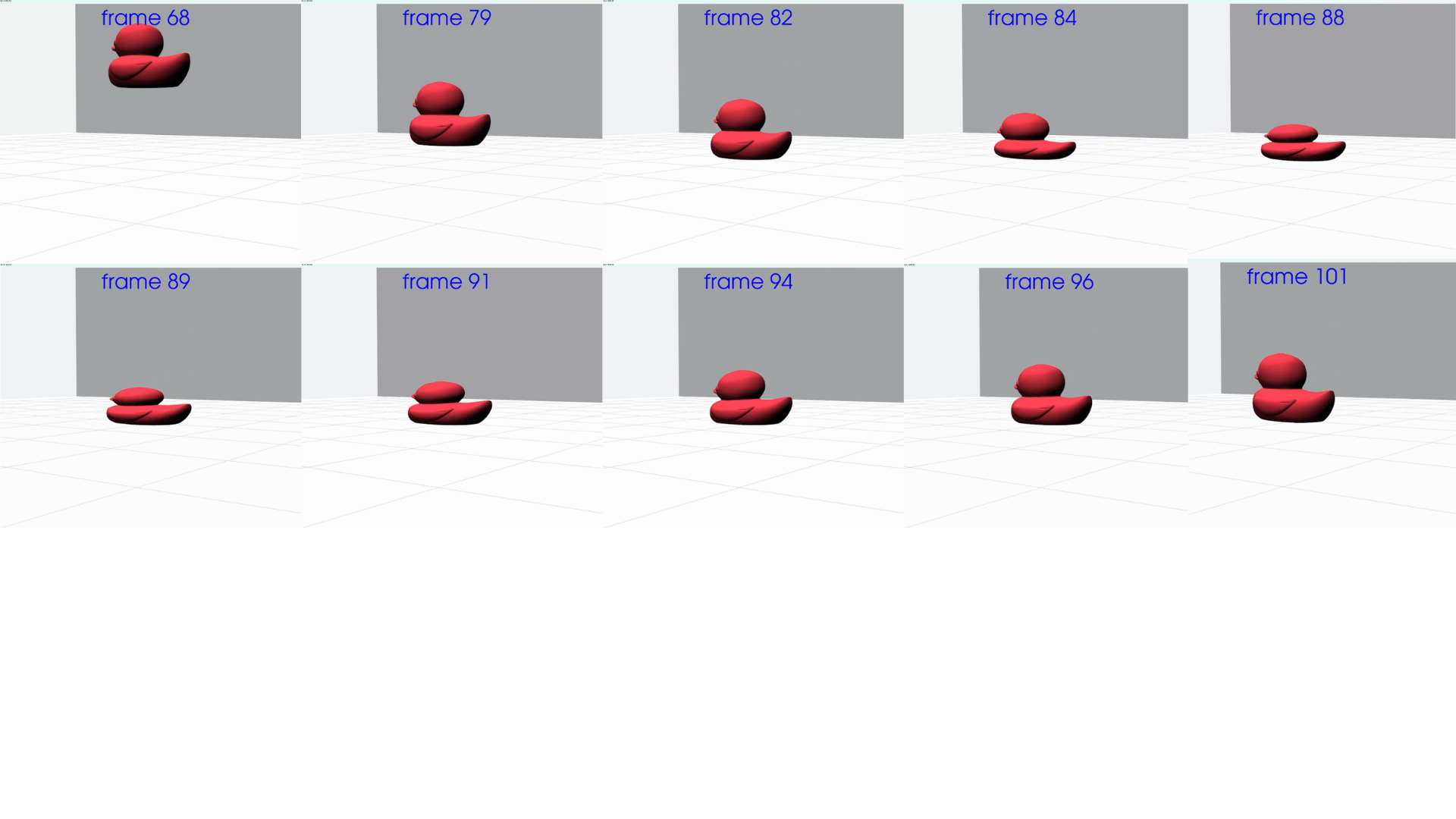}
    
    \vspace{0.5em}
    \begin{minipage}[t]{0.40\textwidth}
        \centering
        \includegraphics[width=\linewidth]{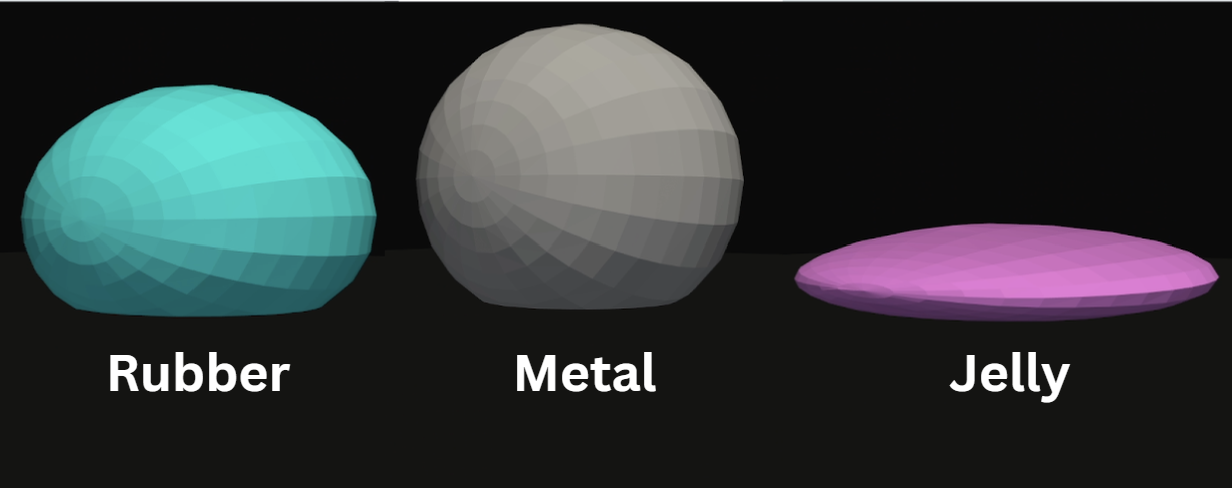}\\[-0.2em]
        \small (a) Representative material responses.
    \end{minipage}
    \hfill
    \begin{minipage}[t]{0.40\textwidth}
        \centering
        \includegraphics[width=\linewidth]{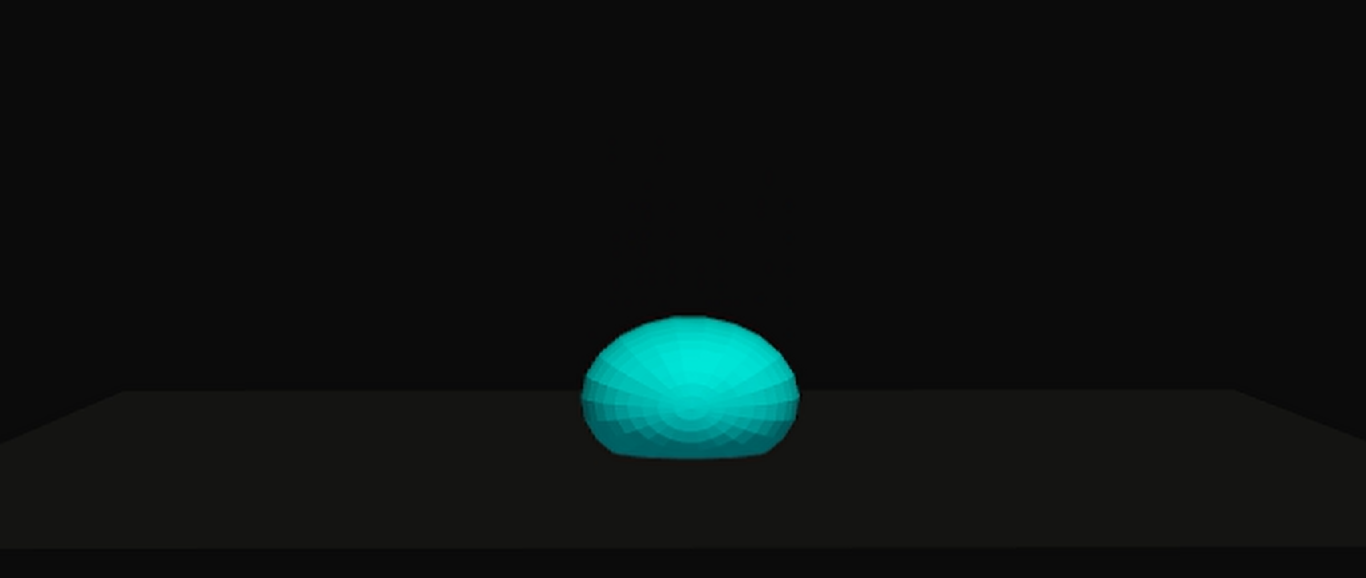}\\[-0.2em]
        \small (b) Covariance ellipsoids expose local strain.
    \end{minipage}
    \caption{Additional reference-entrant diagnostics. Top: a non-spherical
    Gaussian substrate under dynamics. Bottom: material response and local
    covariance deformation from the shared particle representation.}
    \label{fig:duck_sequence}
    \label{fig:four_materials}
    \label{fig:ellipsoids}
\end{figure}

\end{document}